\documentclass[footinbib,twocolumn,showpacs,amsmath,amstex,amssymb,mathfonts,superscriptaddress,prl]{revtex4}
\usepackage{graphicx}
\usepackage{color}
\usepackage{bm}
\usepackage{amsmath}
\usepackage{amssymb}
\usepackage{amsthm}
\usepackage{amsfonts}

\usepackage{bbm}

\begin{document}

\title{Edge Induced Topological Phase Transition of the Quantum Hall state at Half Filling}

\author{Bo Yang}
\affiliation{Division of Physics and Applied Physics, Nanyang Technological University, 637371, Singapore}
\email{yang.bo@ntu.edu.sg}
\affiliation{Complex Systems Group, Institute of High Performance Computing, A*STAR, 138632, Singapore}
\author{Na Jiang} 
\affiliation{Zhejiang Institute of Modern Physics, Zhejiang University, Hangzhou,  310027, P.R. China}
\author{Xin Wan}
\affiliation{Zhejiang Institute of Modern Physics, Zhejiang University, Hangzhou, 310027, P.R. China}
\affiliation{CAS Center for Excellence in Topological Quantum Computation, University of Chinese Academy of Sciences, Beijing, 100190, P.R. China}
\affiliation{Collaborative Innovation Center of Advanced Microstructures, Nanjing University, Nanjing, 210093, P.R. China}
\author{Jie Wang}
\affiliation{Department of Physics, Princeton University, Princeton, New Jersey, 08544, USA}
\author{Zi-Xiang Hu} 
\affiliation{Department of Physics, Chongqing University, Chongqing,  401331, P.R. China,}
\email{zxhu@cqu.edu.cn}
\pacs{73.43.Lp, 71.10.Pm}

\date{\today}
\begin{abstract}
We show that in quantum Hall systems at half-filling, edge potentials alone can drive transitions between the Pfaffian and anti-Pfaffian topological phases. We conjecture this is true in realistic systems even in the presence of bulk interactions that weakly break the particle-hole symmetry. The strong effects of edge potentials could be understood from different topological shifts of competing phases at half filling, manifested on the disk geometry as the variation of orbital numbers at fixed number of particles. In particular, we show analytically particle-hole conjugation of Hamiltonians on the disk is equivalent to the tuning of edge potentials, which allows us to explicitly demonstrate the phase transition numerically.  The importance of edge potentials in various experimental contexts, including the recently discovered particle-hole symmetric phase, is also discussed.
\end{abstract}

\maketitle 

The fractional quantum Hall (FQH) effect, realized with a strong magnetic field perpendicular to a two-dimensional electron system, is a fascinating breeding ground for strongly correlated topological phases\cite{prange}. One particularly interesting family of topological phases is the non-Abelian states with unconventional clustering between electrons\cite{mr,rr}. Quasiparticle excitations from these states have topologically protected internal degrees of freedom, leading to anyonic braiding statistics when one is adiabatically dragged around another\cite{nayak}. The Moore-Read (MR) state in a half-filled Landau level (LL) is one such non-Abelian candidate that has been subjected to intense theoretical and experimental studies\cite{mr,shotnoise,tunnelling,tunnelling2,lin1,lin2, exp1, exp2, banerjee, feldman,feldman2,hu08,hu09} for its topological nature.  

One of the first proposed candidates of the non-Abelian MR states is the Pfaffian (Pf) state, the exact zero energy gapped ground state of a particular three-body Hamiltonian\cite{wen3}. It was soon realized that the Pfaffian state breaks the particle-hole (PH) symmetry at half-filling.  Its PH partner, aptly named as the anti-Pfaffian (APf), is also a potential candidate for the FQH state at half-filling\cite{levin,lee}. More importantly, the Pf and APf states belong to distinct topological phases. Other theoretically proposed topological orders at half-filling include the PH-Pf phase\cite{banerjee,feldman}, $U\left(1\right)\times SU\left(2\right)_2$ order\cite{wen1}, as well as the abelian phases including the $K=8$ state\cite{wen1,wen2}, multicomponent $331$ state and $113$ state (the latter is topologically equivalent to the PH conjugate of the $K=8$ state)\cite{halperin,feldman2,halperin2}. Experimental realization and identification of topological orders at half-filling based on different edge dynamics have been an arduous journey\cite{shotnoise,tunnelling,tunnelling2,lin1,lin2}. Early shot noise\cite{shotnoise} and tunneling experiments\cite{tunnelling,tunnelling2,lin1,lin2} all point to quasiparticles with charge $e/4$, and the tunneling exponents seem to suggest that both APf and 331-like phases\cite{tunnelling2, lin1,lin2} can be realised under different experimental conditions. More recently, thermal Hall conductivity measurements\cite{banerjee} indicate the topological phase realized at half-filling is PH symmetric. The proposed PH symmetric Pfaffian (PH-Pf)~\cite{feldman}, however, has yet to have numerical evidence on being adiabatically connected to incompressible ground states of realistic systems, even for systems with disorder~\cite{halperin2,Lian18, Simon18}.

In the limit of strong magnetic field, the effective Hamiltonian of the FQH systems contains only two-body interactions, which is PH symmetric at half filling for systems without a boundary. %This suggests the ground state has to be PH symmetric on geometries such as torus, and the Pf and APf states have exactly the same variational energy. If Pf or APf is stabilized, we can only realize one of them from either spontaneous PH symmetry breaking, or with explicit symmetry breaking in realistic systems. 
Most of the focuses have been on the possibility of spontaneous PH symmetry breaking induced by effective bulk three-body interactions from LL mixing, in systems with a finite perpendicular magnetic field. Numerical and perturbative calculations show that in general, effective three-body interactions favor the APf phase~\cite{Nayak09, PRL105, PRL106, PRBSimon2013,PRBPeterson2013,PRBSodemann2013}. %The PH symmetry can also be broken with one-body terms in the Hamiltonian. 
An important factor that breaks the PH symmetry in real systems is the detailed one-body potentials near the edge. One of the early works~\cite{hu08} hinted at the possible Pf to APf transition at half filling by tuning the background confining potential. In general, however, the edge effects are ignored in the literature for two reasons: firstly, the edge effects could be unimportant in the thermodynamic limit, where the bulk Hamiltonian should dictate the physics of the topological order from spontaneous symmetry breaking; secondly, most numerical studies of the PH symmetry breaking and the stability of the MR state are performed on sphere or torus geometry, which have smaller finite-size effects. %In the context of mimicking experimental systems, however, the disk geometry should be the most realistic one, especially in cases where edge dynamics may have a leading order effect on the bulk topological properties even in the thermodynamic limit.

In this Letter, we show for a PH symmetric interaction at half-filling, the edge potential could be the dominant factor in breaking the PH symmetry and favoring either Pf or APf phases in the realistic systems. In fact, in the limit of strong magnetic field, tuning the edge potentials alone can drive the incompressible phase at half-filling from Pf to APf and vice versa. Thus if Pf or APf phase is realized at half-filling, explicit PH symmetry breaking by the edge potentials is likely to be more important than spontaneous symmetry breaking in many experimental systems. 

%{\it Topological Shift and edge potentials --}
An intuitive way of understanding the importance of edge potentials is to look at the topological shift of the Pf and APf phase on sphere or disk geometry ($S=-2$ for Pf and $S=+2$ for APf), related to the topologically protected guiding center Hall viscosity\cite{read,haldane}. Given $N_e$ electrons on disk geometry, the model wavefunction can only be realized when the number of orbitals $N_{\text{orb}}=2N_e-2$ (for the Pf state), and $N_{\text{orb}}=2N_e+2$ (for the APf state). %The presence of the shift is due to the coupling of the guiding center spin of the topological fluids to the Hall surface curvature in the bulk(i.e. sphere) or the curvature at the boundary (i.e. disk). 
On the disk with fixed magnetic field, the number of orbitals accessible to electrons can be readily tuned by the edge potentials. Without loss of generality, we assume rotational invariance. Let $H_{\text{1bdy}}=\sum_mV_mc_m^\dagger c_m$ be the one-body local potential term, where $m$ is the orbital index, and $c_m^\dagger, c_m$ are the creation and annihilation operators for electrons. If we take $V_m=0$ for $m\le 2N_e-2$ and $V_m=\infty$ for $m>2N_e-2$, this is equivalent to $S=-2$, giving the Hilbert space that is accessible to the Pf state. However, if we take $V_m=0$ for $m\le 2N_e+2$ and $V_m=\infty$ for $m>2N_e+2$, this gives the Hilbert space for the APf state. Thus the edge potentials for the last four orbitals $ m \in [2N_e-2, 2N_e+2)$ play a significant role: a possible transition from the Pf phase to the APf phase can be induced by lowering $V_m$ in these four orbitals from large values, provided that the interaction supports the Pf/APf phase in the bulk.
\begin{figure}[htb]
\includegraphics[width=\linewidth]{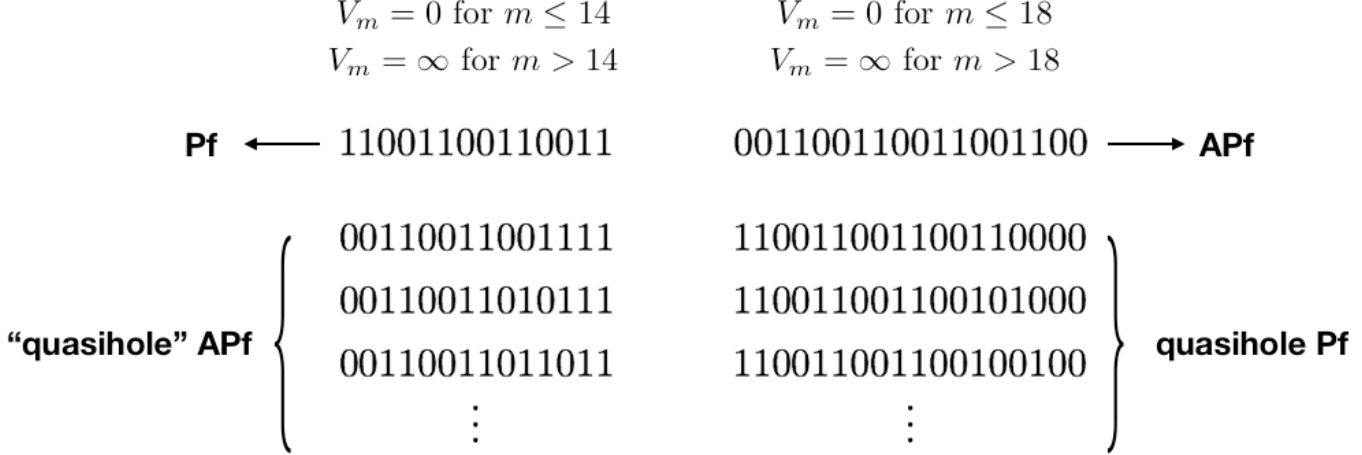}
\caption{An illustration of the possible states (given by their root configurations) for eight electrons, with $2\times 8-2$ orbitals (left) and $2\times 8+2$ orbitals (right). The Hilbert space on the left supports the Pf ground state and various ``quasihole" APf states, while the Hilbert space on the right supports the APf ground state and quasihole Pf states.}
\label{fig1}
\end{figure} 

As shown in Fig.~\ref{fig1}, the Pf (APf) state also needs to compete energetically against all possible ``quasihole" APf (Pf) states. While the two-body interaction by itself favours Pf and APf phase equally, there is a fundamental PH duality not only between Pf and APf ground states, but also between their quasihole excitations. We now carry out detailed analysis on how tuning edge potentials can drive a transition between the Pf and APf phase, where the only source of PH symmetry breaking comes from the presence of the edge.  We assume realistic two-body interactions that can lead to a gapped Pf phase with the right shift ($N_{\text{orb}}=2N_e-2$), supported with extensive numerical evidence\cite{num1,num2}. It can be shown the \emph{same} two-body interaction can lead to a gapped APf phase with the same gap and ground state overlap, by judicious tuning only the edge potentials. 

Starting with the effective microscopic Hamiltonian on the disk as follows:
\begin{eqnarray}\label{masterh}
\mathcal H&=&H_{\text{2bdy}}+H_{\text{1bdy}}\\
H_{\text{2bdy}}&=&\sum_{m,n,m',n'=0}^{N_{\text{orb}}-1}V_{mn}^{m'n'}c_{m'}^\dagger c_{n'}^\dagger c_mc_n\\
H_{\text{1bdy}}&=&\sum_{m=0}^{N_{\text{orb}}-1} V_mc_m^\dagger c_m\label{master1bdy}
\end{eqnarray}
The PH transformation is formally carried out by taking $c_m^\dagger\rightarrow d_m, c_m\rightarrow d_m^\dagger$. Denoting the PH conjugate Hamiltonian as $\mathcal H'$, simple algebra gives us:
\begin{eqnarray}
\mathcal H'&=&\mathcal H+H'_{\text{1bdy}},\quad H'_{\text{1bdy}}=\sum_{m=0}^{N_{\text{orb}}-1}\bar V_mc_m^\dagger c_m\label{ph}\\
\bar V_m&=&4\mathcal V_m-2V_m\label{vm},
\end{eqnarray}
where $\mathcal V_m=\sum_{n=0}^{N_{\text{orb}}-1}V_{mn}^{mn}$ originates from the normal ordering of the PH conjugate of the two-body interactions, physically representing the energy cost of adding an electron to the $m^{\text{th}}$ orbital with a positively charged uniform background. $\mathcal V_m$ depends on $m$ non-trivially especially for $m\sim N_{\text{orb}}$ on disk geometry\cite{footnote1}.

\begin{figure}[htb]
\includegraphics[width=\linewidth]{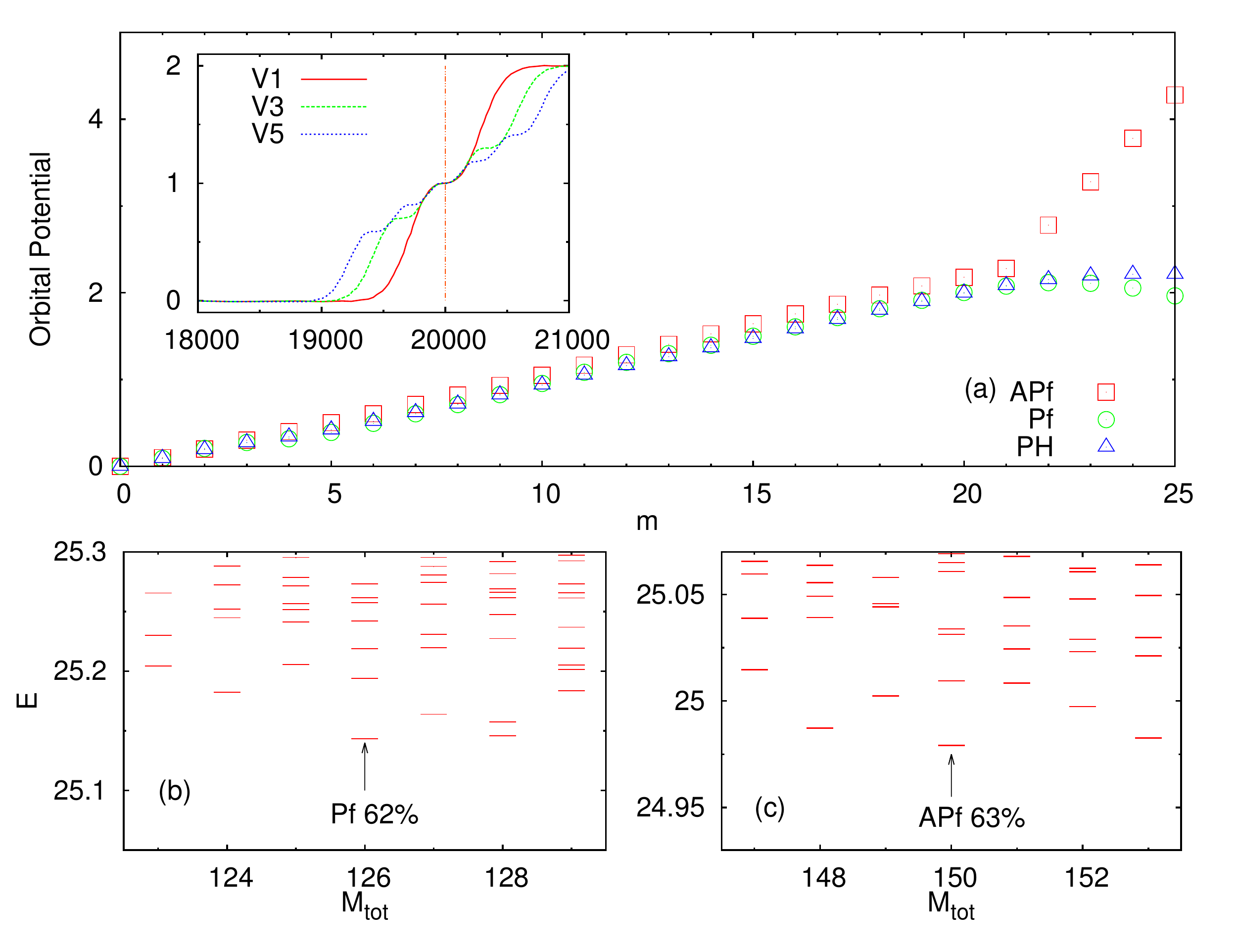}
\caption{a). By keeping the same two-body interaction, tuning the one-body potential leads to the realization of the Pf phase (as seen in b).) and the APf phase (as seen in c).) We also include here the one-body potential that makes the full Hamiltonian PH symmetric. The inset shows one-body potentials from background charge of very large systems (20000 orbitals as indicated by the vertical dotted line), with interactions given by the Haldane pseudopotentials $V_1, V_3, V_5$ respectively. In b) and c) the arrow gives the global ground state, and the percentage gives the overlap with model Pf/APf wave function.}
\label{fig2}
\end{figure}

The important message is that if $\mathcal H$ supports an incompressible topological state (as the global ground state) adiabatically connected to the Pf state with $N_e=N_{\text{orb}}/2+1$, then by adding one-body potential $H'_{\text{1bdy}}$, the resulting $\mathcal H'$ gives the spectrum with the same incompressibility gap at  $N_e=N_{\text{orb}}/2-1$. The global ground state overlap to the APf state is also the same as that of $\mathcal H$ with the Pf state. For small systems, $H'_{\text{1bdy}}$ is non-trivial for orbitals in the bulk (see Fig.~\ref{fig2}). However in the thermodynamic limit, both $\mathcal V_m$ and $V_m$ are only non-uniform near the edge (see Fig.~(\ref{fig2}a) inset). Thus $H'_{\text{1bdy}}$ only significantly affects the edge potentials in real systems, and Eq.(\ref{ph}) could be treated as a rather general analytic proof that tuning the edge potentials alone can allow us to stabilize either Pf or APf phases.

Using finite size disk geometry, it is useful to show explicitly that we can induce a clear phase transition from Pf to APf by just tuning the edge potentials. In particular, starting from Eq.(\ref{masterh}), we fix $H_{\text{2bdy}}$ and only tune $H_{\text{1bdy}}$ to realize different topological phases, by looking at the overlap of the global ground state with model wavefunctions. Realistically the detailed one-body potentials near the edge are always present and can be quite complicated; even the background neutralizing charge contribution depends on the distance of the electron gas from the background, and the local lattice deformation near the edge. As a proof of principle, we tune the edge potentials rather flexibly. This is similar to the theoretical tuning of individual pseudopotentials for the bulk interaction; here tuning the one-body potentials of different orbitals allow us to reveal qualitative physical processes in a transparent manner. In Fig.~\ref{fig2}, we focus on the Hilbert space of $12$ electrons in $26$ orbitals, implicitly assuming that the one-body potentials $V_m$ are very large for $m>25$. The Pf and APf state resides in the total angular momentum sector $M=126$ and $150$ respectively. We show explicitly that by tuning the one-body potentials, especially for those with $m\in[22,25]$, we can have the global ground state at $M=126$ and $M=150$, with overlaps to Pf and APf states at $0.62$ and $0.63$ respectively\cite{footnote2}. The corresponding one-body potentials are shown in Fig.~\ref{fig2}(a). The fixed two-body interaction is a short range interaction consisting of non-zero Haldane pseudopotentials from $V_1$ to $V_9$. We specifically constructed this two-body interaction to enhance the overlap of the global ground states with Pf/APf model wave functions to illustrate the feasibility of such edge potential driven phase transitions. 

%Comparing to the sphere or torus geometry, one disadvantage of the disk geometry is the additional finite size effect for the bulk properties coming from the presence of a non-trivial edge, which should become negligible (to the bulk properties arising from the two-body/three-body interactions) in the thermodynamic limit. There are two ways of compensating for the edge effects to the bulk in the finite systems: adding a linear one-body orbital potential ($V_m$ is linear in $m$), or diagonalising the Hamiltonian in the $L=0$ sector. This is because in the thermodynamic limit, $L$ is a good quantum number.

It is important to understand the competition between bulk LL mixing and edge potentials especially in the thermodynamic limit, for deciding which topological phase will be favoured. If both mechanisms lead to very small PH symmetry breaking, it is generally believed that whether Pf or APf state is experimentally selected depends on spontaneous symmetry breaking. %Formally, we can evaluate the variational energy of the Pf/APf model wavefunction with respect to the PH symmetry breaking terms; the state with the lower variational energy will be selected from spontaneous symmetry breaking. 
In this picture, the bulk interaction from LL mixing should dominate in the thermodynamic limit. %, because the variational energy from the bulk scales with $\sim N_e$, while that from the edge scales with $\sim\sqrt N_e$. 
In real systems, however, effective three-body interactions from LL mixing can be very complicated and fail to prefer Pf over APf (or vice versa), especially for very large magnetic field and LL mixing becomes vanishingly small. In addition, there could be complicated domains of Pf and APf phases\cite{kun} that are relatively small so that the thermodynamic limit does not apply. In these scenario, even if PH symmetry breaking of the edge potentials are subleading to the bulk interaction in terms of the variational energy, they can still play an important role in experimental systems.

In addition, PH symmetry breaking at half filling can be \emph{not} small in experimental systems. Here we show that edge potentials not only explicitly break PH symmetry, but also be a dominant effect in selecting the topological phase at half-filling. For incompressible systems, the bulk gap comes from the two-body interactions and effective three-body interactions from LL mixing, which does not scale with the system size. In general the PH breaking three-body interactions from LL mixing is quite small and suppressed by the strong magnetic field and screening of electron-electron interactions\cite{yang}. On the other hand, the number of orbitals affected by the edge potential scales with $\sqrt N_e$ (see inset of Fig.~(\ref{fig2}a)). As a result, one would expect changes of edge potentials to have significant effects in inducing level crossing, and on determining which topological phase is adiabatically connected to the ground state of realistic systems.
 
\begin{figure}[htb]
\includegraphics[width=\linewidth]{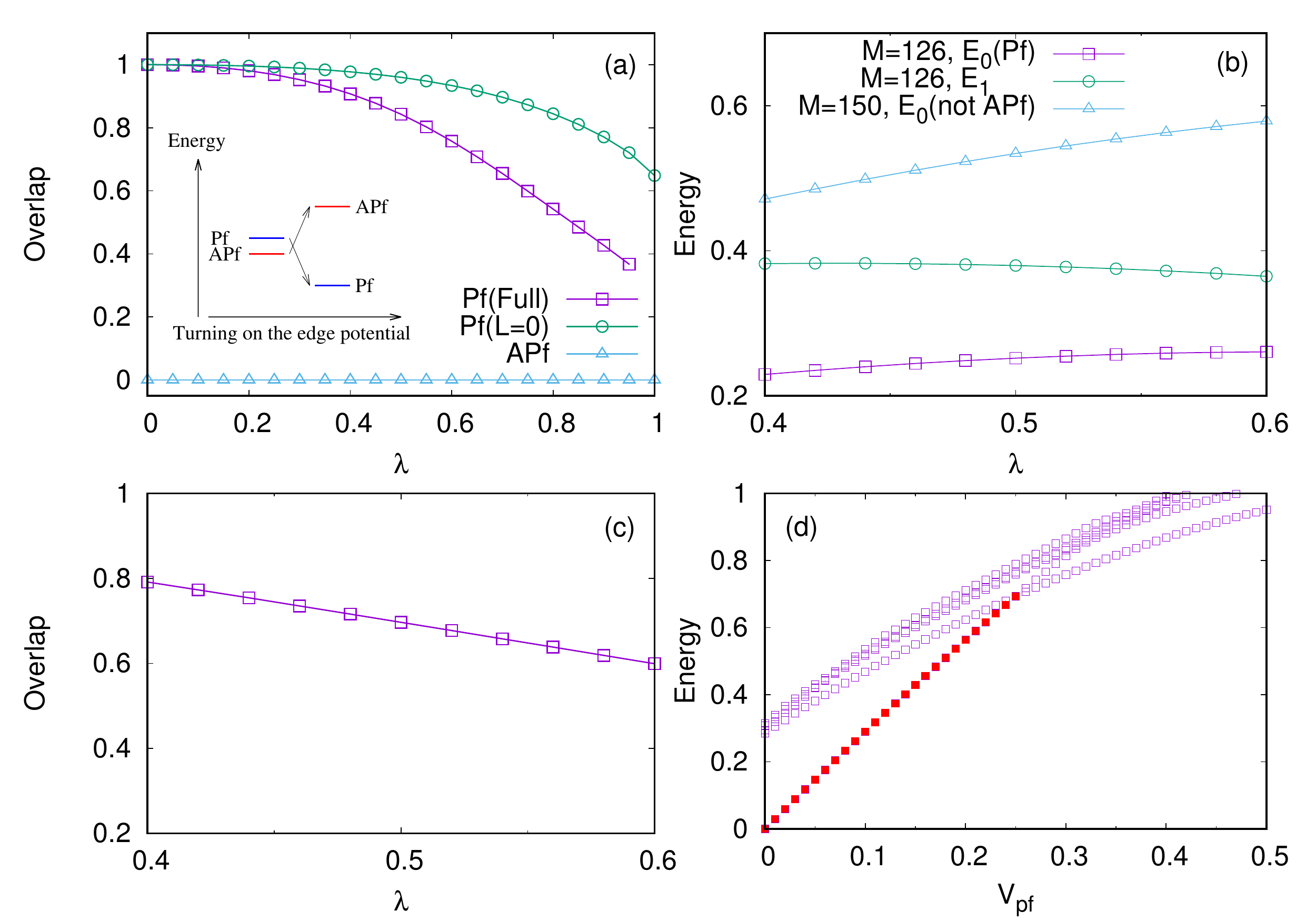}
\caption{a).The overlap of the ground state from Eq.(\ref{hlambda}) with 12 electrons and 22 orbitals, with Pf model wave function (in $M=126$) and with APf (in $M=150$). For b)$\sim$d) we zoom around $\lambda\sim 0.5$, and focus on 12 electrons, 26 orbitals with the last four orbitals imposed with a one-body potential $V_{\text{Pf}}$. The inset of a). schematically shows edge potential induced level crossing, which is explicitly shown in b) with $V_{\text{Pf}}=0.2$; c). the overlap of the ground state in $M=126$ with Pf model wavefunction with $V_{\text{Pf}}=0.2$; d). The energy evolution of low-lying states in $M=150$ with $\lambda=1$. The red part is the wave function with high overlap with APf model wavefunction.}
\label{fig3}
\end{figure}

We look at the extreme case where the bulk interaction is very close to the model Hamiltonian for the APf phase, while the edge potential strongly favors the Pf phase.  We use the following Hamiltonian:
\begin{eqnarray}\label{hlambda}
\mathcal H=\left(1-\lambda\right)\mathcal H_{\text{Pf}}+\lambda\mathcal H_{\text{APf}}
\end{eqnarray}
where $\mathcal H_{\text{Pf}}$ and $\mathcal H_{\text{APf}}$ are model Hamiltonians for Pf and APf states, respectively. In Fig.~\ref{fig3}(a), at $N_{\text{orb}}=2N_e-2$ we clearly see that the edge effects play a dominant role even if the three-body interaction is the \emph{leading} part of the bulk Hamiltonian. One should particularly note that in this Hilbert space, the overlap of the ground state in $M=150$ with the APf model wavefunction is strictly zero. This is \emph{not} a finite-size effect, but is from the squeezing properties of the APf model wave function\cite{jack}, closely related to its topological shift. Only ``quasihole" APf states can be realized with $N_{\text{orb}}=2N_e-2$, and even very close to APf model Hamiltonian ($\lambda\sim 1$), the overlap of $M=126$ ground state with Pf model wave function is very high, indicating those ``quasihole" APf states are not energetically competitive in this angular momentum sector.
At $N_{\text{orb}}=2N_e+2$, we can also add a positive potential $V_{\text{pf}}$ to the last four orbitals, to model an edge potential profile that favors the Pf state. A clear level crossing in $M=150$ subspace is observed at $V_{\text{pf}}\sim0.25$ even with $\lambda=1$ (Fig.~\ref{fig3}(d)). 

The more realistic case is with $\lambda\sim 0.5$, in which PH symmetry breaking by three-body interactions (presumably from LL mixing) is small. In the regime $V_{\text{pf}}>0.25$, small three-body interaction that favors APf phase is inconsequential: the ground state in $M=126$ is manifestly in the Pf phase, while the ground state overlap with APf state in $M=150$ is close to zero. We can also see that the ground state in $M=126$ maintains high overlap with Pf model wavefunction even for relatively low $V_{\text{Pf}}$ (see Fig.~\ref{fig3}(c)), and its energy is significantly lower than the ground state in $M=150$. %(note the state associated with APf is even higher in energy in this sector). 
Thus while the bulk interaction leads to the ground state favoring the APf phase with $\lambda>0.5$, the presence of the edge readily induces level crossing and opens up a larger incompressibility gap for the Pf phase, as schematically illustrated in the inset of Fig.~\ref{fig3}(a).

We now discuss the experimental relevances of the edge potentials at half-filling. Following the recent experiment, it was suggested\cite{banerjee,feldman} the PH-Pf phase could be stabilized by disorder, with the claim that all past experiments detected the PH-Pf phase\cite{feldman}. Even if that is the case, Pf and APf are still well-defined topological phases in the clean limit, and in principle can be realized in experiments. Our work suggests that in the clean limit, realizing Pf or APf phase can strongly depend on the edge potentials in the experimental samples. In particular, a sharp edge potential generally favors the Pf phase (see Fig.(\ref{fig2})), which could be realized despite the bulk interaction (e.g. from LL mixing in Galilean invariant systems\cite{yang}) weakly favoring the APf phase (see Fig.(\ref{fig3})). For smooth edge potentials, APf is generally favored; but it is worth pointing out that the notion of smoothness/sharpness should be considered with respect to the magnetic length $l_B$ defining the width of Landau level orbitals. For the same edge potential profile, with stronger magnetic field the potential varies more smoothly from one orbital to another.

Another possible scenario is that different topological phases have been realized in different experimental settings. It was also previously suggested\cite{lin2} that by tuning the confinement potential (one of the main contributions to the edge potential), the APf phase was detected at weak confinement, while a 331-like phase was observed at strong confinement instead of the Pf phase. One should note that the 331 state has the same topological shift as the Pf state, thus the edge potential favoring the latter also tends to favor the former. Numerical evidence overwhelmingly favors Pf when only bulk interaction is considered\cite{num1,num2,num3,num4,num5}. Most experimental evidence also favours spin polarised states, though a bit less conclusive\cite{willet,csathy,pan,pinczuk} with both spin polarized\cite{exp1,mansour,pan2,stern2,diode} and unpolarized\cite{exp2} possibilities reported. This is understandable, as the issue of spin polarisation likely depends on many experimental parameters and could be different for different experiments or samples.
%Apart from the hypothesis that all incompressible phases measured in the experiments are particle-hole symmetric and stablised by disorder\cite{feldman}, it is also possible that in all past experiment, bulk interactions strongly prefers APf because of strong effects of LL mixing. The latter is unlikely in most realistic samples especially when long range Coulomb interaction is softened (e.g. by finite thickness)\cite{yang}, but it is difficult to rule out. 
One possibility for the absence of Pf is that when strong confinement potential favours Pf over APf in the experiments, the spin unpolarised 331 state is also favoured and thus may compete energetically\cite{footnote}, further reducing the incompressibility gap. One can go to strong magnetic field in the experiments to suppress the spin unpolarised state, but this also results in smaller $l_B$ so that the edge potential profile is effectively more smooth (as given by the dependence of $V_m$ on $m$ in Eq.(\ref{master1bdy})), leading to the APf phase being observed instead. %Thus to experimentally realise the Pf phase, one needs to fabricate edge potentials that are effectively sharp enough even at very large magnetic field.% It is also worth mentioning while detailed edge potential in real space is generally fixed when the sample is fabricated, its effective smoothness in orbital space can still be tuned by varying the strength of the magnetic field. In addition, since to fix the filling factor the electron density also needs to be tuned in most cases with a back-gate voltage, tuning the effective edge potential profiles could be complex but potentially useful in experiments, and is worth further investigations.

%We present general and analytical arguments with detailed numerical computations to show that edge potentials can play an important role in determining the topological orders at half-filling. The analysis should apply to cases where competing topological phases occur at the same filling factor but with different topological shifts. 
{\it Conclusions --} We present general and analytical arguments with detailed numerical computations to show that edge potentials can play an important role in determining the topological orders at half-filling of FQH systems. The analysis should apply to cases where competing topological phases occur at the same filling factor but with different topological shifts. For systems with an edge, there is a bulk-edge correspondence between the bulk guiding center spin and the edge topological shift, and the latter strongly couples to the detailed one-body potentials at the edge. While most attentions in the past focused on bulk interactions, we show explicitly for the competition between Pf and APf phases that the edge potentials can be the dominating effects in principle, which cannot be ignored in experiments especially since the effective edge potential also depends on the strength of the magnetic field.

We conjecture the detailed edge potentials could be important for the stabilisation of the recently proposed PH-Pf phase. PH conjugation on systems with a boundary involves the addition of non-trivial edge potentials (see Eq.(\ref{ph},\ref{vm})), and given any bulk interaction there is a unique edge potential profile for the system to be PH symmetric (see Fig.(\ref{fig2}a)). We thus expect the stability of the PH symmetric topological phase can be strongly affected not only by the bulk dynamics, but also the details at the edge. The PH-Pf is conjectured to be stabilised by disorders in the bulk\cite{feldman,zhuwei}, and our work suggests the interplay between bulk interactions, edge potentials, as well as correlated disorders \emph{both in the bulk and near the edge} could be essential in studying PH-Pf, though unfortunately detailed numerical analysis with disorder on disk geometry is hampered by the finite size limitation.

\begin{acknowledgments}
{\sl Acknowledgements.} We thank X. Lin and M. Shayegan for useful discussions. This work is supported by the NTU grant for Nanyang Assistant Professorship. Z-X. Hu is supported by National Natural Science Foundation of China Grants No. 11674041, No. 91630205, No. 11847301  and Chongqing Research Program of Basic Research and Frontier Technology Grant No. cstc2017jcyjAX0084. X. Wan is supported by the 973 Program under Project No.  2015CB921101, the National Natural Science Foundation of China Grant No. 11674282 and the Strategic Priority Research Program of Chinese Academy of Sciences, Grant No. XDB28000000. 
\end{acknowledgments}


\begin{thebibliography}{99}
\bibitem{footnote1} On sphere or torus geometry, if the LL is fully occupied, $\mathcal V_m$ is independent of $m$ due to translational or rotational symmetry, and is thus trivial.
\bibitem{footnote} For spin unpolarized 331 state, its particle-hole conjugate is at filling factor $7/2$, so it does not compete with the APf phase at weak confinement.
\bibitem{footnote2} Such overlaps are considered very high on the disk geometry, as the presence of the edge mixes different total angular momentum sectors in a finite system. Also see Ref.\cite{hu08}. We can obtain higher overlaps (similar to those on the sphere) by restricting the Hilbert space to the zero center of mass angular momentum sector (so the Hilbert space size is the same as the ones on the sphere), though this will limit the largest system size we can access numerically.
\end{thebibliography}
\end{document}